\documentclass[twoside,twocolumn,9pt]{article}
\usepackage{extsizes}
\usepackage[super,sort&compress,comma]{natbib}
\usepackage[version=3]{mhchem}
\usepackage[left=1.5cm, right=1.5cm, top=1.785cm, bottom=2.0cm]{geometry}
\usepackage{balance}
\usepackage{widetext}
\usepackage{times,mathptmx}
\usepackage{sectsty}
\usepackage{graphicx}
\usepackage{lastpage}
\usepackage[format=plain,justification=raggedright,singlelinecheck=false,font={stretch=1.125,small,sf},labelfont=bf,labelsep=space]{caption}
\usepackage{float}
\usepackage{fancyhdr}
\usepackage{fnpos}
\usepackage[english]{babel}
\usepackage{array}
\usepackage{droidsans}
\usepackage{charter}
\usepackage[T1]{fontenc}
\usepackage[usenames,dvipsnames]{xcolor}
\usepackage{setspace}
\usepackage[compact]{titlesec}
\usepackage{bm}

\usepackage{epstopdf}

\definecolor{cream}{RGB}{222,217,201}

\begin{document}

\pagestyle{fancy}
\thispagestyle{plain}
\fancypagestyle{plain}{

\renewcommand{\headrulewidth}{0pt}
}

\makeFNbottom
\makeatletter
\renewcommand\LARGE{\@setfontsize\LARGE{15pt}{17}}
\renewcommand\Large{\@setfontsize\Large{12pt}{14}}
\renewcommand\large{\@setfontsize\large{10pt}{12}}
\renewcommand\footnotesize{\@setfontsize\footnotesize{7pt}{10}}
\makeatother

\renewcommand{\thefootnote}{\fnsymbol{footnote}}
\renewcommand\footnoterule{\vspace*{1pt}%
\color{cream}\hrule width 3.5in height 0.4pt \color{black}\vspace*{5pt}}
\setcounter{secnumdepth}{5}

\makeatletter
\renewcommand\@biblabel[1]{#1}
\renewcommand\@makefntext[1]%
{\noindent\makebox[0pt][r]{\@thefnmark\,}#1}
\makeatother
\renewcommand{\figurename}{\small{Fig.}~}
\sectionfont{\sffamily\Large}
\subsectionfont{\normalsize}
\subsubsectionfont{\bf}
\setstretch{1.125} 
\setlength{\skip\footins}{0.8cm}
\setlength{\footnotesep}{0.25cm}
\setlength{\jot}{10pt}
\titlespacing*{\section}{0pt}{4pt}{4pt}
\titlespacing*{\subsection}{0pt}{15pt}{1pt}

\fancyfoot{}
\fancyfoot[RO]{\footnotesize{\sffamily{1--\pageref{LastPage} ~\textbar  \hspace{2pt}\thepage}}}
\fancyfoot[LE]{\footnotesize{\sffamily{\thepage~\textbar\hspace{3.45cm} 1--\pageref{LastPage}}}}
\fancyhead{}
\renewcommand{\headrulewidth}{0pt}
\renewcommand{\footrulewidth}{0pt}
\setlength{\arrayrulewidth}{1pt}
\setlength{\columnsep}{6.5mm}
\setlength\bibsep{1pt}

\makeatletter
\newlength{\figrulesep}
\setlength{\figrulesep}{0.5\textfloatsep}

\newcommand{\topfigrule}{\vspace*{-1pt}%
\noindent{\color{cream}\rule[-\figrulesep]{\columnwidth}{1.5pt}} }

\newcommand{\botfigrule}{\vspace*{-2pt}%
\noindent{\color{cream}\rule[\figrulesep]{\columnwidth}{1.5pt}} }

\newcommand{\dblfigrule}{\vspace*{-1pt}%
\noindent{\color{cream}\rule[-\figrulesep]{\textwidth}{1.5pt}} }

\makeatother

\twocolumn[
  \begin{@twocolumnfalse}
\vspace{3cm}
\sffamily
\begin{tabular}{m{4.5cm} p{13.5cm} }

 & \noindent\LARGE{\textbf{Biasing the ferronematic -- a new way to detect weak magnetic field}} \\
\vspace{0.3cm} & \vspace{0.3cm} \\

 & \noindent\large{Nat\'{a}lia Toma\v{s}ovi\v{c}ov\'{a},$^{\ast}$\textit{$^{a}$} Jozef Kov\'{a}\v{c},\textit{$^{a}$} Yuriy Raikher,\textit{$^{b,c}$} N\'{a}ndor \'{E}ber,\textit{$^{d}$} Tibor T\'{o}th-Katona, \textit{$^{d}$} Veronika Gdovinov\'{a},\textit{$^{a}$}  Jan Jadzyn,\textit{$^{e}$} Richard Pin\v{c}\'{a}k\textit{$^{a}$} and Peter Kop\v{c}ansk\'{y}\textit{$^{a}$}} \\

& \noindent\normalsize{Magnetic properties of a ferronematic, i.e., nematic liquid crystal doped with magnetic nanoparticles in low volume concentration are studied, with the focus on the ac magnetic susceptibility. A weak dc bias magnetic field (units of Oe) applied to the ferronematic in its isotropic phase increases the ac magnetic susceptibility considerably. Passage of the
isotropic-to-nematic phase transition resets this enhancement irreversibly (unless the dc bias field is applied again in the isotropic phase). These experimental findings pave a way to application possibilities, such as low magnetic field sensors, or basic logical elements for information storage.} \\

\end{tabular}

 \end{@twocolumnfalse} \vspace{0.6cm}
]


	
\renewcommand*\rmdefault{bch}\normalfont\upshape
\rmfamily
\section*{}
\vspace{-1cm}	

\footnotetext{\textit{$^{a}$~Institute of Experimental Physics, Slovak Academy of Sciences, Watsonov\'{a} 47, 04001, Ko\v{s}ice, Slovakia.  Tel: +421 55 792 2208; E-mail: nhudak@saske.sk}}
\footnotetext{\textit{$^{b}$~Institute of Continuous Media Mechanics, Russian Academy of Sciences, Ural Branch, Perm, 614013, Russia.}}
\footnotetext{\textit{$^{c}$~Ural Federal University, Ekaterinburg, 620083, Russia.}}
\footnotetext{\textit{$^{d}$~Institute for Solid State Physics and Optics, Wigner Research Center for Physics, Hungarian Academy of Sciences, H-1525, Budapest, P.O. Box 49, Hungary.}}
\footnotetext{\textit{$^{e}$~Institute of Molecular Physics, Polish Academy of Sciences, 60179, Poznan, Poland.}}

\section{\label{sec:1} Introduction}

Liquid crystal (LC) research was boosted up dominantly in 1970s by
the liquid crystal display (LCD) technology, and was primarily
focusing on the design, synthesis and characterisation of the LC
materials, as well as on the development of new LCD modes, in order to
fulfill the requirements of the newborn, rapidly expanding LCD
industry. By now, the commercial success of the LCDs has moved
the research and development in this direction mostly
into industrial laboratories \cite{NPG2009}. In the
meantime, the academic LC research has been shifted towards the search
for novel smart functional materials applicable in diverse other fields
such as micro-, nano-, biotechnology, medicine, polymer and
colloid science, photonics, etc.
\cite{Stannarius2009,Lagerwall2012}.

In this respect, the research of LC colloidal systems (various
micro-, or nanoparticles dispersed in LCs) offer a wide range of
possibilities. The use of nematic LCs as the colloidal matrix is
especially of great promise, primarily because nematics provide a
unique opportunity to generate, transform, and control topological
defect (TD) structures (point, line, or sheet disclinations)
\cite{deGennes1993}. TDs in general strongly interact with the
embedded micro-, or nanoparticles, and that can be exploited for
efficient trapping and control of the particles. The TD mediated
self-assembly of particles
\cite{Musevic2006,Ravnik2007,Ognysta2008,Wang2016} offers possible
applications in photonics (e.g., 3D photonic crystals
\cite{Joannopoulos1997}, metamaterials \cite{Pendry2000}), while
the control of TDs may be used e.g., for the guided transport of
microfluidic cargo \cite{Sengupta2014}.

Magnetic nanoparticles (MNPs) dispersed in nematic LCs -- the so
called ferronematics (FNs) -- are the practical manifestation of
the idea put forward by Brochard and de Gennes \cite{BrGe_JP_70}
who suggested that doping liquid crystals with fine magnetic
particles, even in a very low concentration, might significantly
enhance their response to magnetic fields. After the first
implementation of ferronematics \cite{Chen1983}, numerous
experimental works have been done on FNs, which relate to, but are
not limited to: the response of FNs to low magnetic fields
\cite{PoBu_SM_11,BuNe_SM_11,ToTi_PRE_13,MeLi_Nat_13}, the type and
strength of the anchoring at the LC--nanoparticle interface
\cite{KoKo_JMMM_06,ToKo_PT_06,KoTo_PRE_08}, the role of the
functionalization of the nanoparticles \cite{MaKu_JAP_09}, the
magnetic field induced shift of the phase transition temperature
\cite{KoTo_IEEE_11}, the shear flow in FNs \cite{MaZa_JMMM_08},
the dynamics of FNs in magnetic field \cite{BaFi_PRE_94}. Results
of all these efforts evidence that doping the nematic matrices
with a small amount of MNPs affects the properties of the
composite materials considerably, and therefore, nowadays FNs in
the form of stable nematogenic magnetic suspensions are considered
as promising materials due to their high magnetic sensitivity
resulting from a subtle orientational coupling between the
ferromagnetic nanoparticles and mesogenic molecules.

In the same time, to our knowledge, studies on the basic magnetic,
especially magnetodynamic properties of FNs are rather scarce. To
advance in this direction, hereby we investigate the magnetic
susceptibility of FNs in response to a probing alternating
magnetic field. We find an unusual behaviour of the magnetic
susceptibility while passing through the phase transition of the
FN, sensing the prior presence of a small (few Oe) magnetic field,
and providing a concept for potential future applications as
sensors, or logical gates in micro- and nanodevices. We also
provide a phenomenological explanation of the experimental
results, which is related to defect-mediated aggregation of MNPs.

\section{\label{sec:2} Experimental}
\begin{figure}[h]
\centering
\includegraphics[height=5cm]{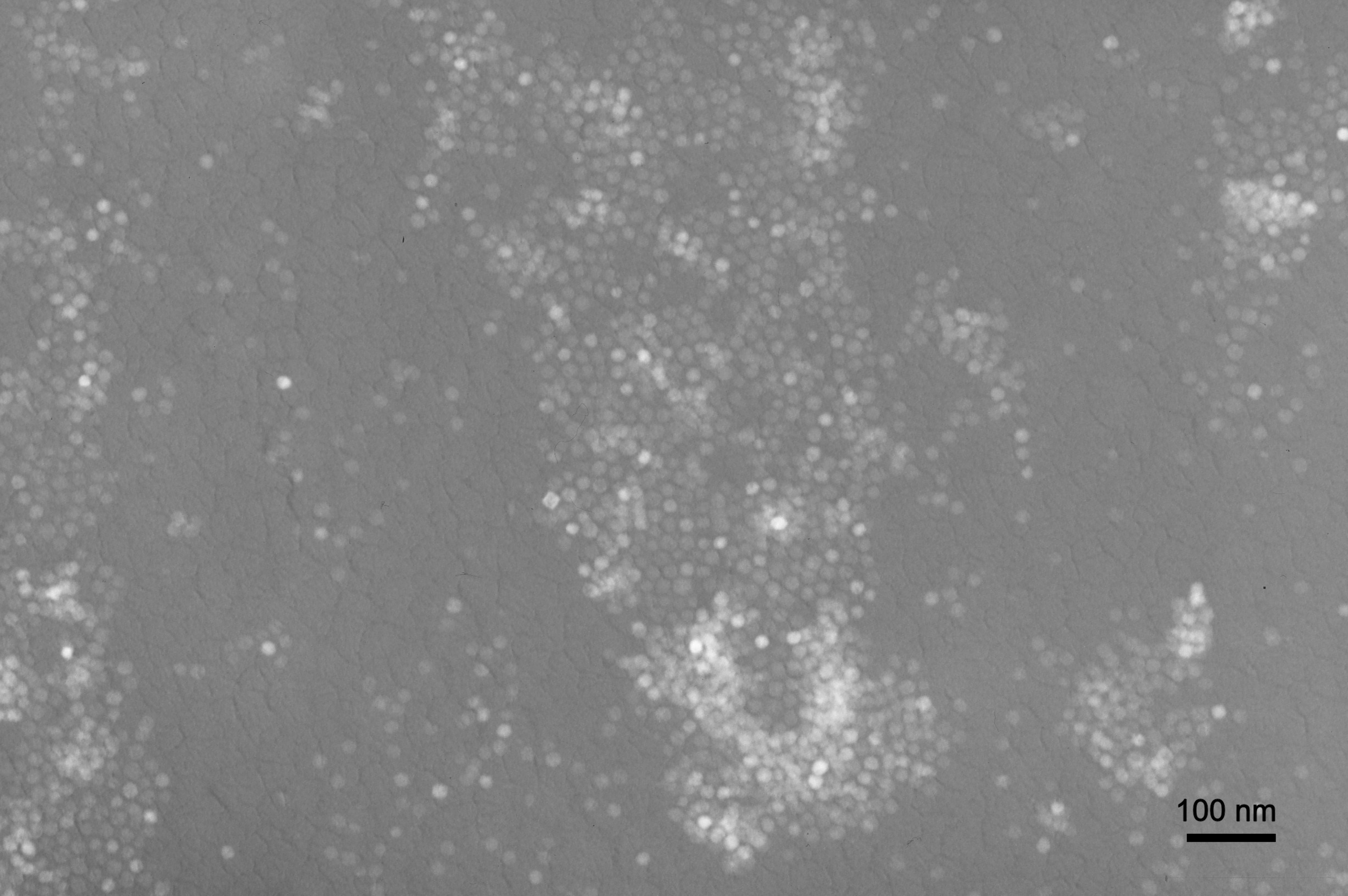}
\includegraphics[height=7cm]{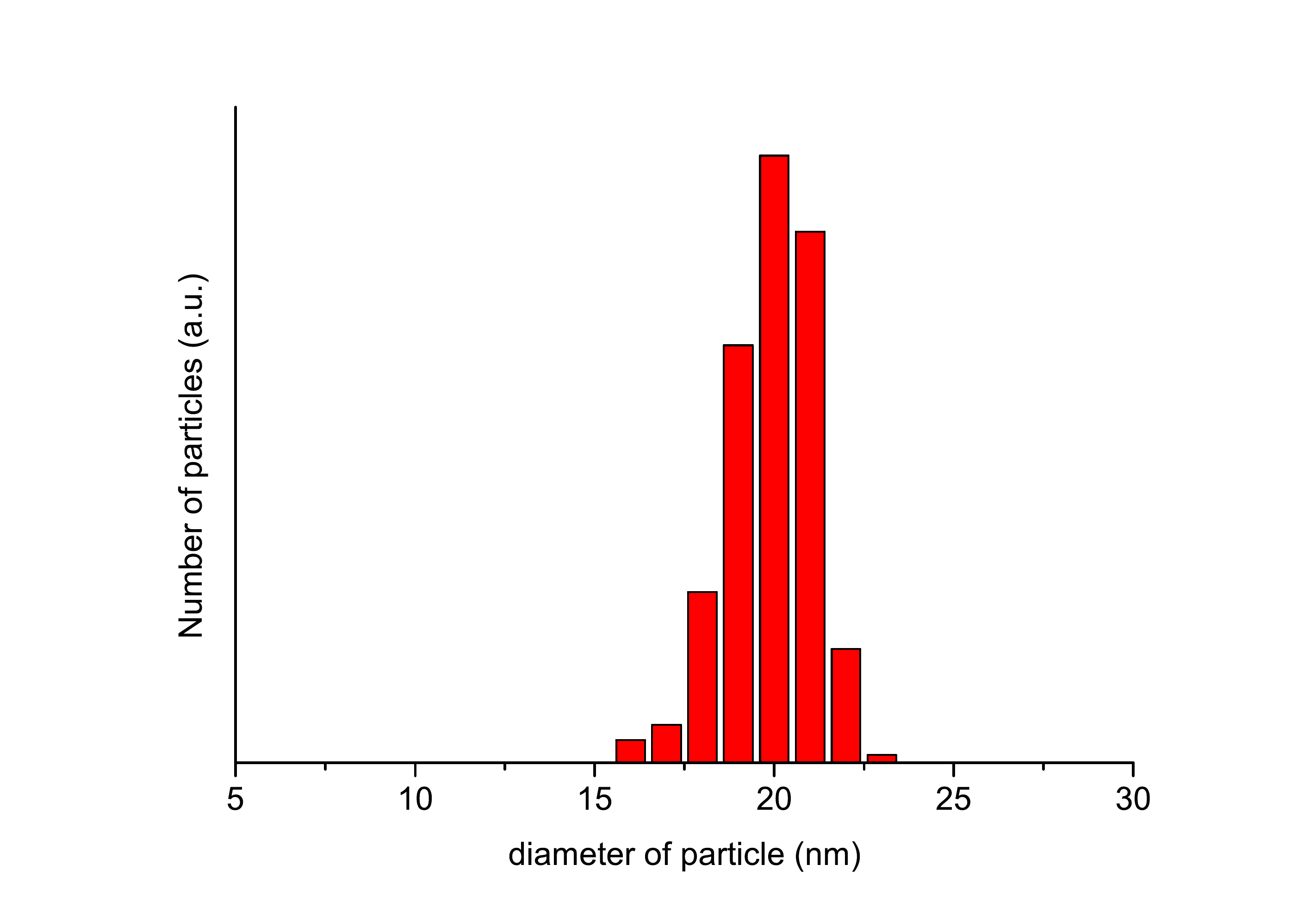}
\caption{(Color online) TEM image of magnetic particles and size distribution.}
\label{sd}
\end{figure}

Measurements were performed in a FN sample based on the calamitic thermotropic liquid crystal n-hexylcyanobiphenyl (6CB)
\cite{GrHa_EL_73,CzCz_Natfor_01}. The temperature $T_{\mathrm{I-N}}$ of the isotropic (I) to nematic (N) phase transition of neat 6CB is 302\,K.
This liquid-crystalline matrix was doped with spherical grains of Fe$_3$O$_4$ (magnetite) of mean diameter
$d=20$\,nm.  {\bf Fig.\ \ref{sd} shows TEM image and size distribution of magnetic particles.}    The volume concentration of the solid phase being $\phi=10^{-4}$. The magnetic nanoparticles were coated with
oleic acid  (providing an appropriate strong interaction of mesogenic molecules with surface of the nanoparticles)  and dissolved in chloroform.
This suspension was admixed to the liquid crystal, and then the solvent was let to
evaporate. For magnetic measurements, both undoped and doped 6CB were filled into cylindrical capsules 2.5\,mm in diameter and
6.5\,mm in length. The magnetic properties were measured with a SQUID magnetometer (Quantum Design MPMS 5XL) in a
magnetic field directed along the cylinder axis of the capsules.

\section{\label{sec:3} Results and discussion}
\begin{figure}[h]
\centering
\includegraphics[height=7cm]{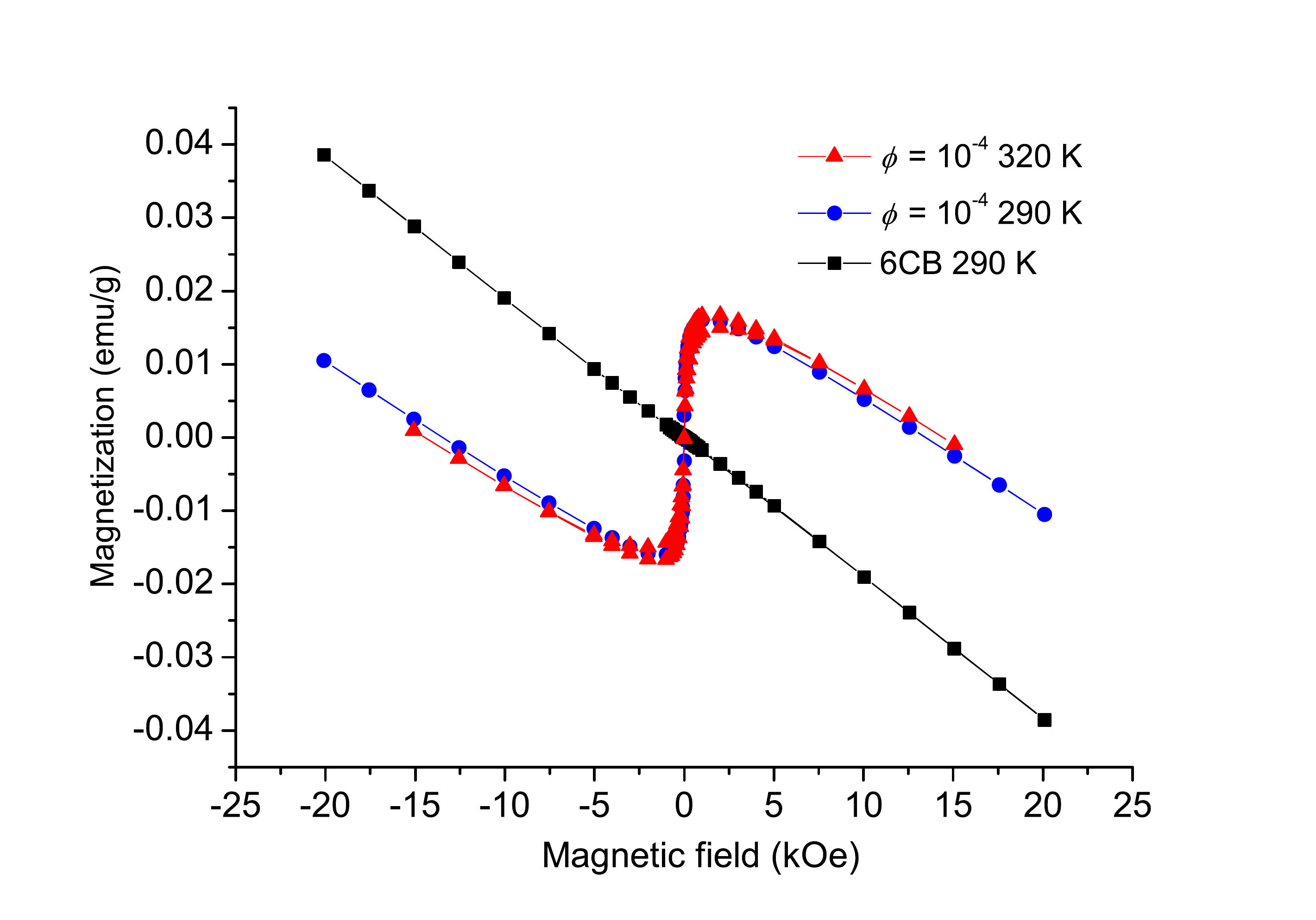}
\caption{(Color online) Magnetization curves of undoped 6CB in nematic phase (squares) and of 6CB-based
FN in nematic and isotropic phases (circles and triangles, respectively).}
\label{fig:01}
\end{figure}

In Fig.\ \ref{fig:01} The magnetization curves of both undoped (neat) and doped 6CB are presented in the nematic phase (at
$T=290$\,K) and, additionally, for the doped sample in isotropic phase (at $T=320$\,K).
The neat 6CB exhibits usual diamagnetic behavior, while the FN composite at low magnetic field behaves as a superparamagnet displaying no hysteresis.
The diamagnetic features of the host matrix become dominating only at higher magnetic fields. The shape of the magnetization curves of the doped sample is
the same in both the isotropic and nematic phases indicating that the quasi-static magnetic properties are independent of the type of fluid phase of the host material.

The dynamic susceptibility $\chi$ of the prepared samples is measured in the same experimental geometry. An ac magnetic field of 1\,Oe is applied at the frequency of $f=650$\,Hz.
To measure the temperature dependence of $\chi$, the samples are first heated to 320\,K (isotropic phase) and then slowly cooled down
(with the rate of 0.5\,K/min) to the nematic phase; after that they are slowly heated up again to 320\,K. The sample is thermally stabilized at each temperature
for 3\,min before performing the susceptibility measurement.

\begin{figure}[h]
\centering
\includegraphics[height=7cm]{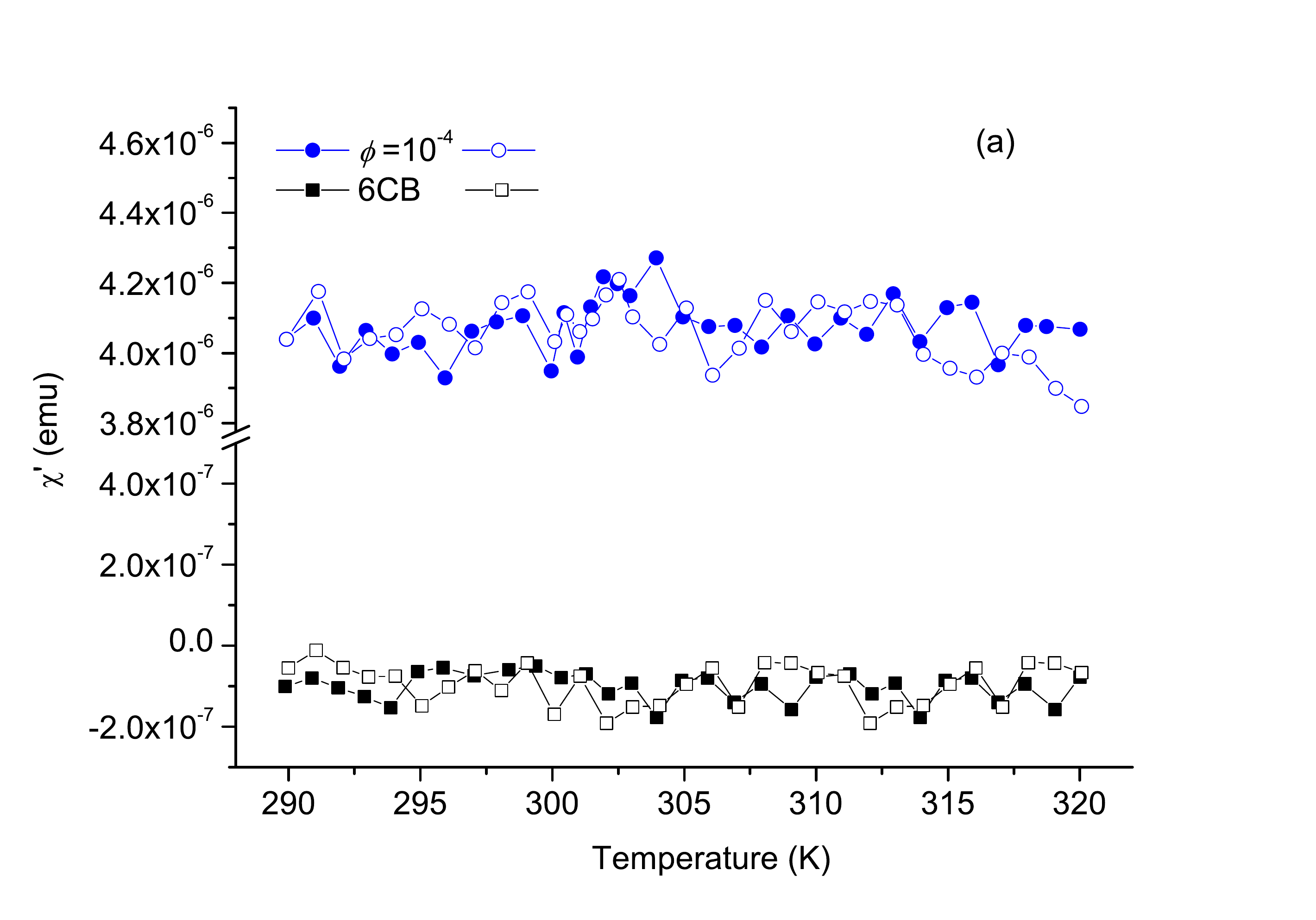} \\
\includegraphics[height=7cm]{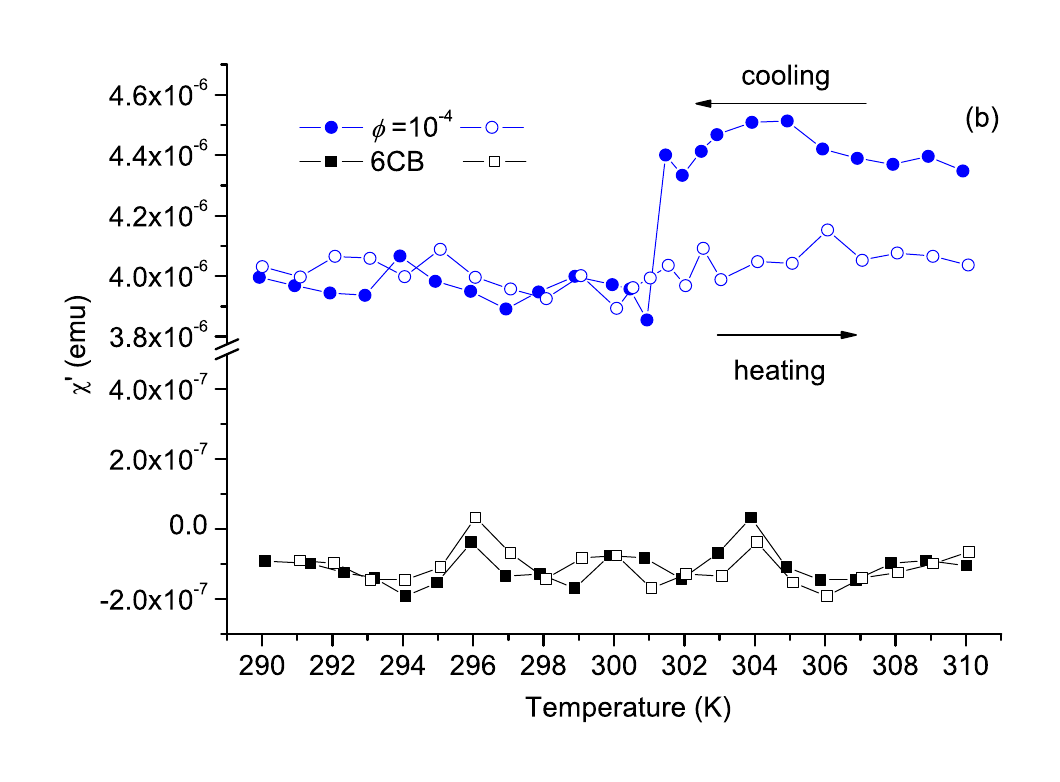}%
\caption{\label{fig:02} (Color online) Temperature dependence of the real part $\chi^\prime$ of ac susceptibility of the
undoped 6CB (squares) and 6CB-based FN (circles) in the cooling-heating cycle; (a) without prior application of magnetic field,
(b) with prior application of a dc magnetic field $H_{\mathrm{dc}}=20$\,kOe. Solid symbols are for cooling, open ones for the heating part of the cycle.}
\end{figure}

Figure \ref{fig:02}a shows the temperature dependence of the real component $\chi^\prime$ for both undoped and doped 6CB.
As it should be, the ac susceptibility of the neat 6CB is small and negative. In contrast to that, the response of the FN sample is positive and much higher
than that of the undoped 6CB in agreement with the magnetization curves shown in Fig.\ \ref{fig:01}.
Notably, for both samples the temperature dependence of $\chi^\prime$ is rather weak.
Moreover, no change is detected when passing through the I-N phase transition, either on cooling or on heating. The susceptibility values,
\begin{equation}\label{eq:chi_IN}
\chi^\prime_{\mathrm{I}}\simeq\chi^\prime_{\mathrm{N}}\simeq4.0\cdot10^{-6}\,\mathrm{emu},
\end{equation}
of the FN are well reproducible through several cooling-heating cycles.

Figure \ref{fig:02}b presents the result of a similar experiment, except that prior to the same cooling-heating cycle, the samples were subjected to a
dc magnetic field $H_{\mathrm{dc}}=20$\,kOe in the isotropic phase viz.\ at 320\,K. In case of a neat 6CB, the magnetic field does not alter the temperature dependence:
the curve in Fig.\ \ref{fig:02}b reproduces the one in Fig.\ \ref{fig:02}a.
A surprising fact, however, is that under this condition the ac susceptibility of the FN in the isotropic phase,
\begin{equation}\label{eq:chi_I_H}
\chi^\prime_{\mathrm{I}_H}\simeq 4.4\cdot10^{-6}\,\mathrm{emu}
\end{equation}
is higher than $\chi^\prime_{\mathrm{I}}$ in Fig.\ \ref{fig:02}a.
Moreover, on cooling, an abrupt change of $\chi^\prime$ occurs at the temperature corresponding to the isotropic-to-nematic phase transition in the neat 6CB.
Namely, $\chi^\prime$ drops from $\chi^\prime_{\mathrm{I}_H}$ back to $\chi^\prime_{\mathrm{N}}\simeq\chi^\prime_{\mathrm{I}}$.
Upon heating, the ac susceptibility remains at that value, i.e., no change occurs on passing the N-I phase transition.
During subsequent cooling-heating cycles (without application of a dc magnetic field), $\chi^{\prime}$ remains unaltered.
The higher $\chi^\prime_{\mathrm{I}_H}$ value could be restored, however, by re-applying $H_{\mathrm{dc}}$ in the isotropic phase.
We note that switching of $H_{\mathrm{dc}}$ on and off in the nematic phase and then heating back to above $T_{\mathrm{I-N}}$ does not affect the susceptibility of the ferronematic.

\begin{figure}[h]
\centering
\includegraphics[height=7cm]{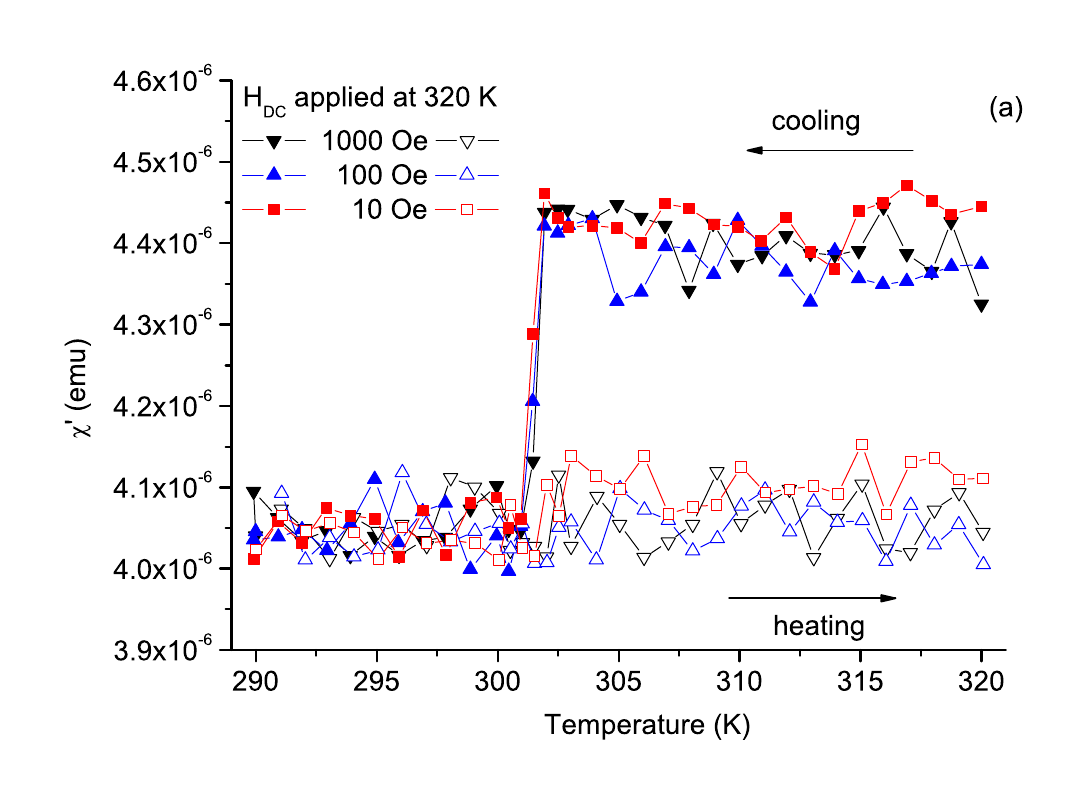} \\%
\includegraphics[height=7cm]{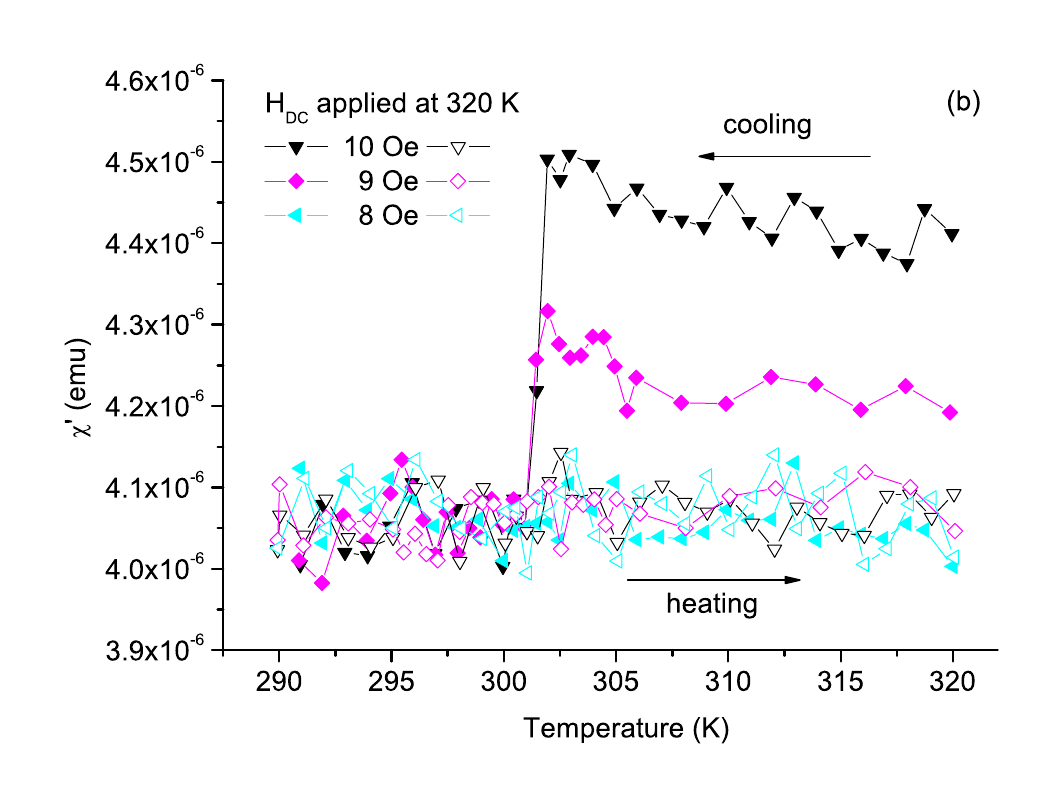}
\caption{\label{fig:03} (Color online) Temperature dependence of the real part $\chi^\prime$ of ac susceptibility of 6CB-based FN measured in a cooling-heating cycle after
applying and switching off a dc magnetic field $H_{\mathrm{dc}}$
(a) higher than 10\,Oe and (b) lower than 10\,Oe. Solid symbols are for cooling, open ones for the subsequent heating.}
\end{figure}

Figure \ref{fig:03} presents $\chi^{\prime}(T)$ under the same conditions as in Fig.\ \ref{fig:02}b, except that the
$H_{\mathrm{dc}}$ values are lower. As Fig.\ \ref{fig:03}a shows, the same higher value of isotropic ac susceptibility
$\chi^\prime_{\mathrm{I}_H}$ and hence the jump
\begin{equation}\label{Delta_chi}
\Delta \chi^{\prime} =
\chi^\prime_{\mathrm{I}_H}-\chi^\prime_{\mathrm{I}}\simeq4.0\cdot10^{-7}\,\mathrm{emu}
\end{equation}
at $T_{\mathrm{I-N}}$ upon cooling may be induced by dc magnetic fields of different magnitudes. One finds that $\Delta
\chi^{\prime}$ ranging about 10\%{} of $\chi^\prime_{\mathrm{I}}$, is independent of $H_{\mathrm{dc}}$ in a wide range:
$10\,\mathrm{Oe}\leq H_{\mathrm{dc}}\leq 20$\,kOe. Figure \ref{fig:03}b evidences, however, that there exists quite a low ``critical''
value $H_{\mathrm{dc}}\approx 9$\,Oe below which the induction effect disappears.

From the above-presented observations we infer the following.
First, a dc magnetic field applied in the isotropic phase alters the structure of the FN in such a way that its ac susceptibility increases.
Second, the increase of the susceptibility is due to MNPs, since it is absent in the neat liquid crystal.
Third, the magnetic field-induced structural changes are removed when the composite enters the nematic phase.
As a step towards understanding this behavior, let us estimate the main parameters which govern the response of the FN to the probing ac and bias dc magnetic fields.

The FN system in question is an ensemble of magnetite MNPs coated with oleic acid and suspended in the matrix of 6CB.
For the estimations below we take the mean diameter of the particles as $d=20$\,nm, their volume concentration $\phi=10^{-4}$,
and the temperature of the I-N phase transition in the matrix as 300\,K.

As is ubiquitous for magnetite nanodispersions, the magnetic anisotropy of the particles deduced from experiments is higher than the crystallographic one
inherent to bulk crystals \cite{GoBe_JAP_03}. For estimates, we set the volume energy density of the magnetic anisotropy to $K=2\cdot10^5$\,erg/cm$^3$ as in
Ref.\ \cite{FoMi_JACS_07}. Relating the anisotropy energy of a particle, $KV$, to thermal energy $k_{\mathrm{B}}T$, where $V=d^3\pi/6$ is the volume of an MNP,
$k_{\mathrm{B}}$ the Boltzmann constant and $T$ the absolute temperature, one arrives at the nondimensional parameter
\begin{equation}\label{eq:sigma}
\sigma=\frac{KV}{k_{\mathrm{B}}T}=\frac{\pi K d^3}{6k_{\mathrm{B}}T}\simeq20.
\end{equation}
Such a value evidences that the particles are virtually free of the N{\'e}el superparamagnetism.
In other words, each particle behaves as a nanosize permanent magnet, and its magnetic moment $\bm{\mu}$ is tightly stuck to a certain internal axis.
Thus, vector $\bm{\mu}$ cannot rotate in response to a magnetic excitation (ac field) otherwise than together with the particle body.

For a single-domain particle, the magnitude of its magnetic moment is defined as $\mu=M_sV$, where $M_s$ is saturation
magnetization of the ferromagnet. Setting for nanodispersed magnetite $M_s=400$\,emu/cm$^3$ (see e.g., \cite{GoBe_JAP_03}), one obtains
\begin{equation}\label{eq:mu}
\mu=\frac{\pi M_s d^3}{6}\simeq1.6\cdot10^{-15}\,\mathrm{emu}\sim1.7\cdot10^5\cdot\mu_{\mathrm{B}},
\end{equation}
where $\mu_{\mathrm{B}}$ is Bohr magneton.

The magnetic coupling of the particles is described by the pairwise dipole-dipole potential. The reference intensity of this
interaction is given by the nondimensional parameter
\begin{equation}\label{eq:lambda}
\lambda=\frac{\mu^2}{k_{\mathrm{B}}Td^3}\simeq8.
\end{equation}
With allowance for the results of former studies of aggregation effects in magnetic fluids \cite{MoPs_JMMM_87}, one concludes that this value is quite high.
Therefore, in the situation under study, particle aggregation is highly probable.
In other words, for such a system a thermodynamically stable state implies the presence of certain amount of multi-particle aggregates. This, in turn,
means that even if in the initial state all the particles are positioned apart from one another, they tend to aggregate.
This process might be rather slow but could be accelerated if the interparticle interaction is enhanced by aligning the magnetic moments with an external magnetic field.
As the considered ensemble is rather dilute, when undergoing aggregation from the state of isolated particles, the most probable entities to emerge are pair aggregates
(``dimers'') \cite{GoBe_JAP_03,SzVa_PRE_13}.

The theoretical estimates for the magnetization curve in the isotropic phase of the system are as follows.
The magnetization $M$ of a dilute ensemble of isolated particles suspended in a fluid is
\begin{equation}\label{eq:M(H)}
M(H)=\phi M_sL(\xi), \qquad \xi=\mu H/k_{\mathrm{B}}T,
\end{equation}
where $L(\xi)=\coth \xi-1/\xi$  is the Langevin function, and $\xi$ is the Langevin argument which relates the Zeeman energy
(the energy of interaction of the particle magnetic moment with the applied field) to thermal energy.

The static (equilibrium) initial susceptibility $\chi_0$ of the system that describes the initial slope of the magnetization curve is defined as
\begin{equation}\label{eq:chi_0_def}
\chi_0=\frac{M(H)}{H} \qquad \mathrm{in\ the\ limit~} H\rightarrow 0,
\end{equation}
As the Langevin function at $\xi\rightarrow 0$ scales as $L(\xi)\approx\xi/3$, one has \cite{RaSh_ACP_94}:
\begin{equation}\label{eq:chi_0}
\chi_0=\frac{\phi M_s^2 V}{3 k_{\mathrm{B}} T}\simeq5\cdot10^{-4}\,\mathrm{emu}.
\end{equation}

In experiments, the limit $\xi\rightarrow 0$ is often not accessible, e.g., when recording the magnetization curve shown in Fig.\ \ref{fig:01}, the lowest magnetic field is 50\,Oe
since SQUID is not reliable below that value. Given that, we use $\chi_{50}=M(H)/H$ measured at $H=50$\,Oe to approximate the static susceptibility.
Keeping in mind that for our system this magnetic field amounts to $\xi\simeq 2$, so that function $L(\xi)$, although being far from saturation,
displays pronounced nonlinearity, we surmise that the true $\chi_0$ is about 20\%{} higher than $\chi_{50}$.

Evaluation of the data shown in Fig.\ \ref{fig:01} sets the experimental value to

\begin{equation}\label{eq:chi_50}
\chi_{50}\simeq3.7\cdot10^{-6}\,\mathrm{emu}, \quad \mathrm{that\enspace implies}\quad\chi_0\simeq4.6\cdot10^{-6}\,\mathrm{emu}.
\end{equation}

This value turns out to be about 100 times lower than the
estimation in Eq.\,(\ref{eq:chi_0}). One cause of this discrepancy
could be that the actual volume concentration of magnetite in the
ferronematic is lower than the nominal one determined from the
synthesis conditions. Another, seemingly more relevant, cause is
that the model of well-dispersed, isolated MNPs [used in deriving
Eq.\,(\ref{eq:chi_0})] fails for real samples. In above it is
noted, see Eq.\,(\ref{eq:lambda}), that the interparticle
interactions are strong enough to provoke aggregation, i.e.,
formation of multi-particle clusters. In the initial field-free
isotropic state, such clusters are assumed to take configurations
where the magnetic flux is nearly closed. Due to that, their
effective magnetic moments are much smaller than the sum of those
for the same number of single particles. In result, the
contribution of multi-particle clusters to the magnetic
susceptibility is quite low, except for the case of strong
magnetic fields, which are not attainable in dynamic measurements.

Dealing with the dynamic susceptibility requires the analysis of its frequency dispersion properties.
>From general considerations, one infers that under a weak ac magnetic field only overdamped, i.e., forced overdamped oscillations are possible, which are rendered by a Debye-type formula
\begin{equation}\label{eq:Debye}
\chi(\omega)=\frac{\chi_0}{1-i\omega\tau}=\chi^\prime+i\chi^{\prime\prime},
\quad \chi^\prime(\omega)=\frac{\chi_0}{1+\omega^2\tau^2},   \quad
\chi^{\prime\prime}(\omega)=\frac{\chi_0\cdot\omega\tau}{1+\omega^2\tau^2};
\end{equation}
where $\omega=2\pi f$ and $\tau$ is the reference response time. Eq.\ (\ref{eq:Debye}) shows that maximum of dissipation occurs the excitation frequency $f^\ast=2\pi/\tau$.
In the range $f>f^\ast$ the ac susceptibility becomes considerably smaller than the static one ($\chi^{\prime}\ll\chi_0$), while for $f\ll f^\ast$ one has $\chi^\prime\approx \chi_0$.

\begin{figure}[h]
\centering
\includegraphics[height=7cm]{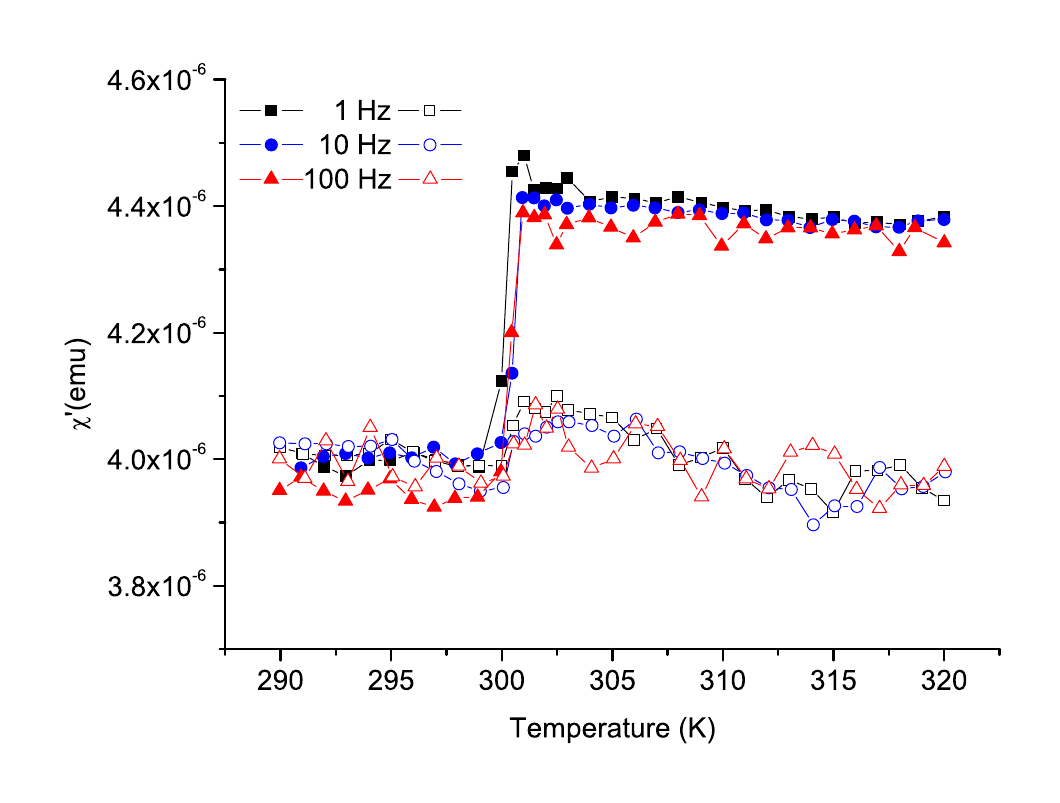}%
\caption{\label{fremu} (Color online) Temperature dependence of the real part $\chi^\prime$ of ac susceptibility of 6CB-based ferronematic measured at different frequencies in a
cooling-heating cycle after applying and switching off a dc magnetic field $H_{\mathrm{dc}}$. Solid symbols are for the cooling, open ones for the subsequent heating.}
\end{figure}

The experiments presented in Figs.\ \ref{fig:02} and \ref{fig:03} are performed at frequency of $f=650$\,Hz. Figure \ref{fremu}
shows the temperature dependence of $\chi^\prime$ measured under the same conditions, but at various (lower) exciting frequencies.
From Figs.\ \ref{fig:02}b, \ref{fig:03}a and \ref{fremu} it is seen that the susceptibilities $\chi^\prime_{\mathrm{I}}$, $\chi^\prime_{\mathrm{I}_H}$, $\chi^\prime_{\mathrm{N}}$ and,
thus, the magnitude of the jump $\Delta \chi^{\prime}$ at the I-N phase transition are practically independent of the frequency within almost 3 decades.
>From this observation one concludes that, on the one hand, the observed jump in $\chi^\prime$ is not due to the frequency dispersion and, on the other hand,
the characteristic time of the system is $\tau\ll2.5\cdot10^{-4}$\,s.

In the adopted model, the reference time of the system under weak ac excitation should be taken as the Brownian rotary diffusion time $\tau_{\mathrm{B}}=3\eta V/kT$
of a spherical MNP of volume $V$ suspended in a fluid with the dynamic viscosity $\eta$ \cite{AbAl_JMMM_93}.
Upon entering the nematic phase, the viscosity of the matrix changes \cite{ViLa_LB_95} and becomes anisotropic;
moreover, elastic torques acting on the particle might emerge as well.
Therefore, one anticipates a change of the characteristic time $\tau$ of the system below $T_{\mathrm{I-N}}$.
However, the above-presented results evidence that $\tau$, despite its changes at the phase transition,
remains small enough not to induce dispersion in the studied frequency range.
Therefore, one concludes that the behavior of $\tau$ cannot be responsible for the jump in $\chi^\prime$ at $T_{\mathrm{I-N}}$.

The lack of dispersion implies that the measured dynamic susceptibility does not differ much from the quasi-static one.
Comparison of $\chi^\prime_{\mathrm{I}_H}$ in Eq.\,(\ref{eq:chi_I_H}) with the actual static susceptibility obtained from magnetization
measurements, see Eq.\,(\ref{eq:chi_50}), confirms this inference.

\begin{figure}[h]
\centering
\includegraphics[height=7cm]{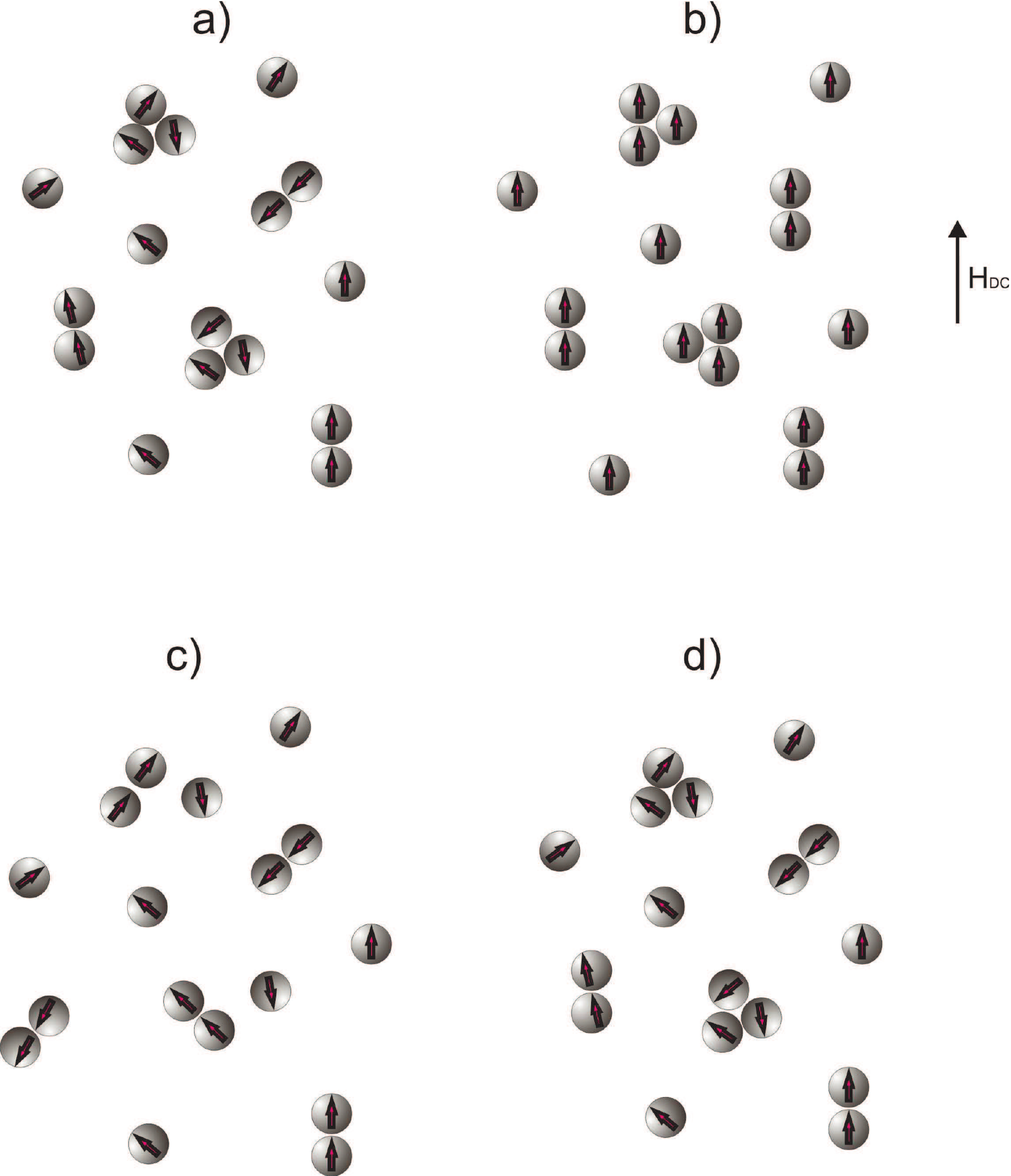}
\caption{\label{fig:model} Schematic representation of the proposed arrangements of the contributing (single MNP or dimer) and
non-contributing (trimer) magnetic particles in the ferronematic: (a) isotropic phase; (b) isotropic phase after applying a magnetic field $H_{\mathrm{dc}}$;
(c) isotropic phase after switching off the magnetic field; (d) nematic phase after cooling down from isotropic phase.}
\end{figure}

Summing up the afore-presented evidence, one sees that a weak dc magnetic field induces an enhancement of the ac susceptibility of a FN,
which effect disappears when passing the I-N phase transition. In our view, the occurrence of this effect we associate with the details of aggregation of MNPs in a nematogenic matrix.
Proceeding to explanation, we note in the isotropic phase of FN the nanoparticles are not ideally dispersed in the host liquid.
As a consequence of Eq.\,(\ref{eq:lambda}), single particles, if they are located close enough, attract each other and strive to form units of two MNPs (dimers),
trimers and higher-order clusters. Once having been formed, those clusters are unbreakable by thermal motion. While trimers and multi-particle clusters assume
configurations with low magnetic moments (nearly closed flux state), the dimers do not.
They rather behave as elongated particles with the magnetic moment along their major axis.
In the isotropic phase, the magnetic moments of single MNPs and of those united in dimers are disordered due to the orientational Brownian motion;
Fig.\ \ref{fig:model}a displays a schematic representation of this state; for simplicity, the clusters with more than three particles are not shown.
The magnetic moments of single MNPs and dimers are modulated  by the probing ac magnetic field yielding $\chi^\prime_{\mathrm{I}}$,
while clusters with near-to-zero magnetic moment virtually do not contribute.
This explains why there is a substantial difference between the estimated $\chi_0$ in Eq.\,(\ref{eq:chi_0}) and the measured static and dynamic susceptibilities
[see Eqs.\,(\ref{eq:chi_50}) and (\ref{eq:chi_IN})].

Application of a strong dc magnetic field in the isotropic phase, aligns the magnetic moments in the direction of $H_{\mathrm{dc}}$, as shown in Fig.\ \ref{fig:model}b.
For triangular trimers, however, the parallel magnetic moments correspond to an energetically unfavored configuration;
it is unstable and prefers to disintegrate into a dimer and a single MNP which repel each other.
Similarly, larger aggregates may also become aligned by strong fields and may loose particles due to repulsion forces.
When $H_{\mathrm{dc}}$ is switched off, the magnetic moments become disoriented by Brownian rotation,
while multi-particle clusters go back to their closed-flux shapes. Due to partial disintegration of trimers and other multi-particle clusters,
the magnetic field-treated isotropic sample (Fig.\ \ref{fig:model}c) contains more particles contributing to $\chi^{\prime}$ (single MNPs and dimers)
and less non-contributing ones (trimers and higher-order clusters) than the untreated FN (Fig.\ \ref{fig:model}a). As a result, a higher ac
susceptibility, $\chi^\prime_{\mathrm{I}_H}$ of Eq.\,(\ref{eq:chi_I_H}), should be detected.

When this system is cooled down and driven through the I-N transition, all the suspended solid entities---single particles as well as clusters of all kinds---induce
nematic structure disclinations around themselves \cite{Musevic2006,Wang2016}. Due to that, on the one hand binding forces between particles
may appear \cite{Ravnik2007,Ognysta2008}, and on the other hand the matrix accumulates extra orientation-elastic energy and strives to get rid of this excess by expelling the particles.
In chemical terms, this means diminution of the particle solubility, which manifests itself as enhanced particle aggregation.
Under those conditions, any single particle ``isolated'' in result of the magnetic field treatment is easily attracted by a dimer (or a larger cluster)
forming a triangular or other aggregate with a near-to-zero net magnetic moment.
Due to that, the number of contributing magnetic moments becomes smaller yielding $\chi^\prime_{\mathrm{N}}$ of Eq.\,(\ref{eq:chi_IN}).
Accordingly, the system in the nematic phase acquires the structure illustrated by Fig.\ \ref{fig:model}d, which is virtually identical to that of Fig.\ \ref{fig:model}a.

When the FN is heated to isotropic state, all the disclinations ``thaw'', and the stimulus for further aggregation disappears.
We remind again that, due to the high value of $\lambda$ in Eq.\,(\ref{eq:lambda}), a once-formed cluster does not break apart under thermal motion.
Heating of the system back to 320--340\,K does not change much in this situation.
Therefore, the system returns to the isotropic state (Fig.\ \ref{fig:model}a) with the same number of clusters, which it has acquired when being in the nematic state.
This implies that $\chi^\prime_{\mathrm{I}}\approx\chi^\prime_{\mathrm{N}}$ as stated by Eq.\,(\ref{eq:chi_IN}).

Therefore, according to the proposed model, during the cooling-heating cycle shown in Fig.\ \ref{fig:02}a the FN passes the sequence of states
(a)~$\rightarrow$~(d)~$\rightarrow$~(a) of Fig.\ \ref{fig:model}.
In case of the experiments presented in Figs.\ \ref{fig:02}b, \ref{fig:03} and \ref{fremu}, the system changes its state following the sequence
(a)~$\rightarrow$~(b)~$\rightarrow$~(c)~$\rightarrow$~(d)~$\rightarrow$~(a) of Fig.\ \ref{fig:model}.

\section{\label{sec:4} Conclusions}
We have demonstrated that a small dc magnetic field (of the order of several Oe)  applied in the isotropic phase modifies the magnetic susceptibility of a ferronematic by about 10\%.
This enhanced value subsists for a long time ($\sim$\,hours) while the sample is kept in the isotropic phase.
Driving it through the isotropic-to-nematic phase transition resets the magnetic susceptibility to the value measured prior to the application of the dc bias field.
After that, the sample could be ``biased'' again by repeated applications of the dc field in the isotropic phase.

The proposed phenomenological explanation associates the discovered effect with the aggregation of nanoparticles in the course of the isotropic-to-nematic phase transition
and their disaggregation under the influence of a dc (bias) magnetic field.
Therefore, it is not surprising that the effect is inherent only to ferronematics and it has no analogues in undoped liquid crystals.

Although here we presented the experimental results for a single concentration of magnetite nanoparticles in a specific liquid-crystalline matrix,
the reported effect of biasing appears to be generic: our preliminary experiments on the composites based on various liquid-crystalline matrices and various ferrite fillers,
reveal the same effect. Given that, we surmise that biasing of ferronematics with a low magnetic field could be useful for future utilization in various micro- and nanodevices e.g., as sensors or logical gates. These applications require fine tuning of the properties of novel types of ferronematic materials, especially targeted for information storage applications.
Such optimization of ferronematics---with respect to the matrix and the type as well as the concentration of nanoparticles---seems to be definitely worth of future studies.

\section{Acknowledgments}
This work was supported by projects VEGA 2/0045/13 and 1/0861/12, the Slovak Research and Development Agency under the contract No.
APVV-0171-10 and APVV-SK-HU-2013-0009, Ministry of Education Agency for Structural Funds of EU in frame of projects 6220120021,
6220120033 and 26110230097, EU FP7 M-era.Net - MACOSYS (Hungarian Scientific Research Fund OTKA NN 110672) and the Hungarian
National Research, Development and Innovation Office grant T\'{E}T\_12\_SK-1-2013-0025. Y.R. acknowledges support of grant
15-12-10003 from Russian Science Foundation. {\bf We are grateful to Ivo V\'{a}vra for TEM image.}

\end{document}